\relax
\documentclass[12pt]{article}
\usepackage{latexsym}
\usepackage[]{}

\begin{document}

\title{Exotic Structures On Magnetic Multilayers}

\author{Feodor V Kusmartsev$^{1,2}$, H S Dhillon$^1$ 
and M D Crapper$^1$}

\date{}

\maketitle

{\bf  $^1$ Department of Physics, Loughborough University, LE11 3TU, UK}

{\bf $^2$ Landau Institute, Moscow, Russia}
\vspace{5mm}

\vspace{4mm}


\begin{abstract}
To characterize the possible magnetic 
structures created on magnetic multilayers 
a model has been formulated and studied.  
The interlayer inhomogeneous  structures found indicate 
either (i) a regular periodic, (ii) a quasiperiodic 
change in the magnetization or (iii) spatially chaotic glass 
states.  
The magnetic structures created depend mainly on the ratio
 of the magnetic anisotropy constant to the exchange constant. 
 With the increase of this ratio the periodic structures first
 transform into the quasiperiodic and then into the chaotic 
glass states. The same tendency arises with the depolarization 
of the magnetic moments of the first layer deposited on the 
substrate. 
\end{abstract}	

\begin{tabbing}

  \= Correspondence to: Professor F Kusmartsev$^{1}¥$, Tel: +44 (0) 1509 223316, \\
  \> Fax: +44 (0) 1509 223986; email: F.KUSMARTSEV@lboro.ac.uk

\end{tabbing}

\vfill
\eject

Modern growth techniques, such as molecular beam epitaxy (MBE) or laser 
ablation, allow magnetic mono- or 
multilayers to be built up.  The magnetic layers produced 
from $Fe$, $Ni$ or $Co$ may be separated by non-magnetic layers, produced, for 
example, from $Cu$.  
In these structures the magnetic films 
are grown one layer at a time.
  In many cases the single layer appears as a single domain, i.e. 
all magnetic moments having a single orientation \cite{Hows}. 
When a second layer is grown on top of 
the first layer the orientation of magnetic moments in the
second layer is not necessarily the same as in the first layer.  Following 
addition of a third layer, the magnetic moments of this layer 
may have yet another orientation.  
The orientation of the magnetic moments is usually 
dictated by a competition between non-uniformity of exchange energy and 
anisotropy energy.  Therefore, the questions arise : 
what kind of magnetic structures (analogous to the Bloch domain wall 
in bulk magnetic samples) may 
be created by the interaction between the monolayers in the film; 
how many types 
of structures can be created; what are the energy costs to create 
such a structure?  The estimation of such energies and their hierarchy will 
indicate the possible temperatures and other conditions of the substrate 
needed for the creation of such structures.

To describe the magnetic structure in a (uniaxially symmetric) magnetic 
multilayer the exchange energy and 
the anisotropy energy have been taken into consideration.  
The exchange energy, $E_{ex}$, favours the 
alignment of the magnetic moments of atoms 
and the magnetic anisotropy energy, $E_{an}$, promotes the alignment 
of the magnetic 
moments along the `easy' axis or `easy' plane depending on what kind 
of (uniaxial) magnetic anisotropy is dominant in the system.

A multilayer film has 
different exchange constants depending on whether the exchange is developed 
in-plane or inter-plane. The nonmagnetic layers 
separating the magnetic layers may also contribute to the 
exchange constants between magnetic layers.  As 
exchange coupling is a short-range effect 
the exchange constant inside each layer is much larger than the constant 
associated with the exchange interaction between the layers.  
Increasing the space between layers decreases the inter-layer exchange coupling 
but not the constant of magnetic anisotropy energy which
is usually related to a long-range spin-spin interaction.  
There are some observations in $Co-Cu$ films that with an 
increase of the interlayer spacing the exchange 
constant strongly decreases while the constant in the anisotropy term 
fluctuates slightly \cite{Back}.

In general the orientation of the magnetic moments depends on two angles 
$(\theta, \phi)$ associated with the in-plane and inter-plane rotation 
of the moments.  However, as the in-plane exchange interaction is much 
stronger than the inter-plane exchange interaction, 
it is easier to cause a defect in the alignment of the magnetic 
moments of different layers than to create a defect inside a single 
plane and so the in-layer moments may be considered as a single domain 
whereas interlayer moments should not be.  Therefore, as a first step, only 
inter-plane inhomogeneous magnetic structures created, 
assuming that all magnetic moments in the same layer align 
homogeneously, are considered.

With this assumption 
the relevant terms of the Hamiltonian associated with the interlayer magnetic 
structure are an interlayer exchange energy and the anisotropy energy.  The 
competition between these two terms determines the interlayer structure 
of the magnetic multilayer film.  
The Euler-Lagrange equation for this model
\begin{equation}
  - x_{n-1}¥ + 2 x_{n}¥ - x_{n+1}¥ + \beta \sin x_{n}¥ = 0  \label{eq1}
\end{equation}

\noindent is a discrete version of the
Sine-Gordon equation where $\beta$ is the ratio of the constant of 
anisotropy energy to the constant of exchange energy and $x_{n}¥$ is 
the orientation of the magnetic moments of the $n^{th}¥$ layer from 
the `easy' axis.  To 
solve this equation we apply the methods of Chaotic Dynamics.  
With such an approach \cite{KusK}, 
instead of solving the Sine-Gordon equation directly, we derive a 
2D discrete map and investigate the trajectories of this map.  
The simple ferromagnetic alignment of the moments is a fixed point of 
(\ref{eq1}) and 
does not depend on the value of $\beta$.  However, 
there are always fluctuations associated with a finite 
number of layers that destroy any such alignment which make it impossible 
for this structure to exist.  
We find three characteristically different
types of trajectories which may be classified as periodic, quasiperiodic and 
chaotic.  These trajectories will correspond to the creation of 
three types of magnetic structures: periodic, quasiperiodic and chaotic, 
respectively.

It is found that the structures created 
depend on both the orientation of the magnetic moments in the 
first layer, $x_{0}¥$ and on the value of $\beta$.
Two types of periodic magnetic structures are 
created at small values of $\beta$.  The first is 
spin density waves frozen in space.  
In the second type of periodic structure the orientation of the 
magnetic moments perform a rotation as we move up through the layers.  
These rotations may have either a positive or a negative sign.  In analogy with 
vortices in superfluid systems one may refer to this structure as a periodic 
structure of spin vortices.  With the increase of the parameter $\beta$ from 
zero there initially arise frozen spin density waves, the period 
of which decreases as 
the value of $\beta$ increases.  Then, at some value of $\beta$, there appear 
spin vortices which are periodically separated throughout 
the multilayer film creating a lattice.  
These spin vortices (see Fig 1) 
occur over a relatively small number of layers while their 
separation is very large.  
The first 35 layers have approximately the same orientation 
and are nearly perpendicular 
to the substrate, $x_{0} \simeq 10^{-5}$.  This region is 
schematically indicated by the three rows of disks immediately above 
the substrate in Fig 1.  The deviation from the vertical axis progressively 
increases for the next 20 layers of the multilayer and spin rotation 
occurs.  
After this rotation there are about 
seventy-five layers with approximately vertically oriented magnetic moments 
and then the next spin vortex occurs.  This structure is periodically 
repeated.

With further increase of the parameter $\beta$ the distance 
between the spin vortices decreases and also begins to fluctuate.  
When this distance becomes equal to the size of the spin vortex 
the quasiperiodicity is broken and some exotic, chaotic structures 
arise.  In such structures, together with spin vortices, there are also
incomplete spin rotations.  Such structures 
arise only at large values of $\beta$ and may be equivalent to spin glasses 
arising in bulk magnetic samples. 

In conclusion, we have found that in thick magnetic films associated 
with magnetic multilayers there may arise 
spin density waves, spin vortex lattices and 
possibly chaotic magnetic structures.

\noindent {\large \bf Figure Caption}

\noindent {\bf Fig 1} The cross section of the magnetic multilayer film 
displaying a spin vortex arising in the 35th - 51st layers
of a quasiperiodic magnetic structure with an approximate period 
of 90 magnetic layers.   

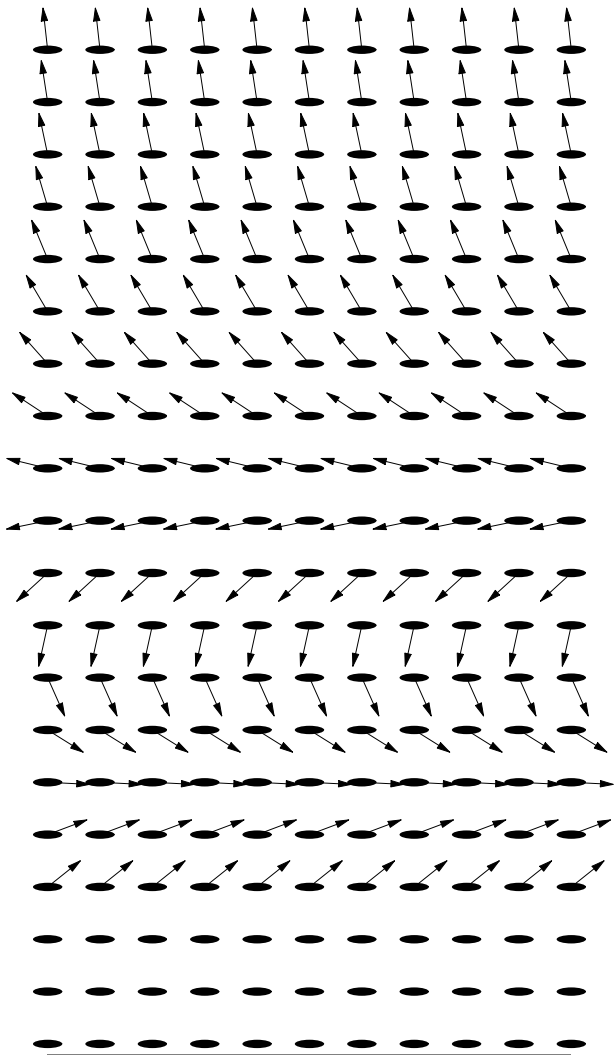
\end{document}